\RequirePackage{fix-cm}
\documentclass[twocolumn,epjc3]{svjour3}  
\usepackage{amsmath}
\smartqed  
\RequirePackage{graphicx}
\journalname{Eur. Phys. J. C}
\newcommand{\pt}{$p_{\rm{T}}$}

\newcommand{\meanpt}{$\langle p_{\rm T} \rangle$}
\newcommand{\cv}{$C$}

\newcommand{\mt}{$m_{\rm{T}}$}
\newcommand{\mT}{$m_{\rm{T}}$}
\newcommand{\meanmt}{$\langle m_{\rm T} \rangle$}

\newcommand{\Tbin}{$T_{\rm bin}$}
\newcommand{\Tevt}{$T_{\rm eff}$~}

\newcommand{\sNN}{$\sqrt {{s_{\rm NN}}}$}

\newcommand{\energy}{$\langle \epsilon  \rangle$}
\newcommand{\temp}{$\langle T \rangle$}

\newcommand{\yphi}{$y$--$\phi$} 

\begin{document}

\title{Characterization of relativistic heavy-ion collisions at the
  Large Hadron Collider through
  temperature fluctuations 
}
\author{Sumit Basu\thanksref{e1,addr1}
            Rupa Chatterjee\thanksref{addr1} 
            Bastanta K. Nandi\thanksref{addr2} 
        \and
            Tapan K. Nayak\thanksref{addr1} 
}

\thankstext{e1}{e-mail: sumit.basu@cern.ch}

\institute{Variable Energy Cyclotron Centre, Kolkata - 700064, India \label{addr1}
           \and
           Indian Institute of Technology Bombay, Mumbai - 400076, India \label{addr2}
}

\date{Received: date / Accepted: date}

\maketitle

\begin{abstract}

We propose to characterize heavy-ion collisions 
at ultra-relativistic energies by using fluctuations of energy density
and temperature. Temperature fluctuations on an event-by-event basis
have been studied both in terms of global temperature of the event,
and locally by constructing fluctuation maps in small phase space bins
in each event.
Global temperature fluctuations provide an estimation of the specific heat of the system.
Local temperature fluctuations of the event may be ascribed to the remnants
of initial energy density fluctuations. Together these two observables give an
insight into the system created in heavy-ion
collisions and its evolution. Event-by-event
hydrodynamic calculations indeed provide adequate
theoretical basis for understanding the origin of the fluctuations.
We demonstrate the feasibility of studying global and local temperature 
fluctuations  at the Large Hadron Collider
 energy 
by the use of AMPT event generator.


\end{abstract}

\maketitle


\section{Introduction}
Heavy-ion collisions at ultra-relativistic energies 
create matter at extreme conditions of energy
density ($\epsilon$) and temperature ($T$), similar to the ones that
existed within a few microseconds after the Big Bang~\cite{heinz}. 
The fireball produced in the collision goes through a rapid
evolution from an early partonic phase of
deconfined quark-gluon plasma (QGP) to
a hadronic phase and ultimately freezing out after a few tens of fm. 
The major goals of colliding heavy-ions at the Relativistic
Heavy Ion Collider (RHIC) and at the Large Hadron Collider (LHC) are to
study the nature of the phase transition and understand the QGP matter
in detail. With the production of large number of particles 
in each of the collisions, it has become possible to extract
thermodynamic quantities on an
event-by-event basis, 
rather than averaging over a sample of events.
Event-by-event fluctuations of $\epsilon$, $T$, mean
transverse momentum, particle 
multiplicity, particle ratios, etc., as well as fluctuations of
conserved quantities within finite detector acceptances have been
proposed to provide dynamical information regarding the evolving 
system~\cite{Sto,steph1,shuryak,steph2,wilk1,wilk2,nayak,jeon-koch,gavai-gupta,borsanyi,Ejiri}.

\begin{figure*}[htbp]
\centering
\includegraphics[width=0.9\textwidth]{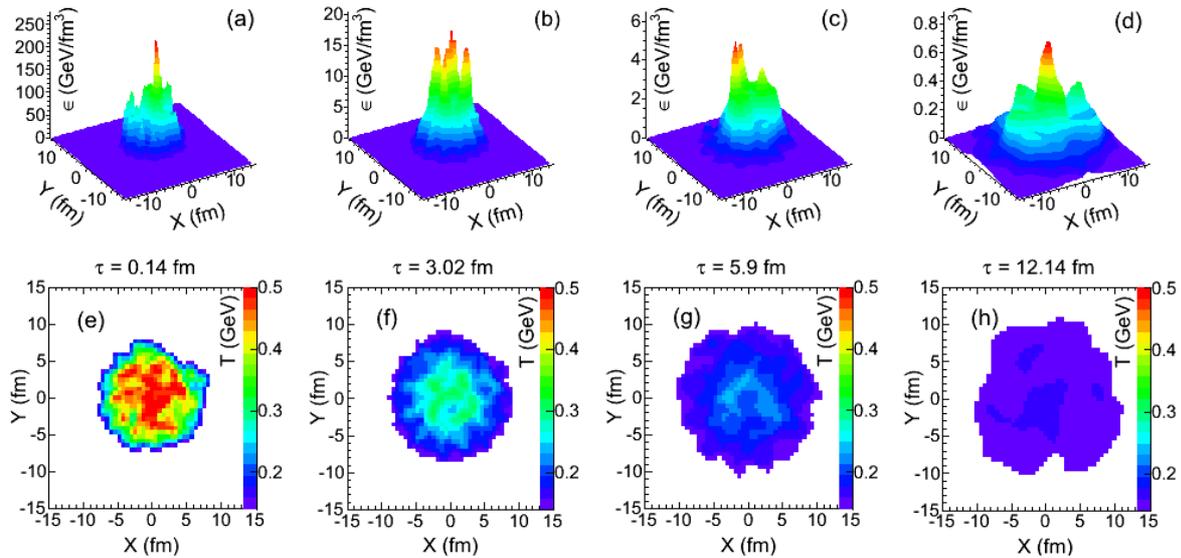} 
\caption
{(Color online). Distributions of energy density (upper panles) and temperature
(lower panels) in the transverse ($X$-$Y$) plane at four proper times 
($\tau$) obtained from hydrodynamic calculations for a single central
Pb-Pb event at \sNN~=2.76 TeV.
}
\label{fig1}
\end{figure*}

In this manuscript, temperature fluctuations have been studied on an
event-by-event basis by estimating global temperature of the event, and local
temperatures in small phase space bins within the event.
Determination of temperature and its fluctuation
for heavy-ion collisions have been possible because of large number of particles
emitted in each event~\cite{alice_energy,alice_centrality}, which is
essential to keep the statistical uncertainties of the measurements 
within a reasonable limit.
Global event to event temperature fluctuations provide the heat
capacity (\cv),
which is an important thermodynamic quantity in characterizing the
system. Lattice QCD calculations predict 
a strong temperature dependence of the heat capacity, 
the nature of which depends on the order of the phase
transition~\cite{shuryak,gavai-gupta-cv}. This study sheds light
on the nature of phase transition at the LHC energy. 
Fluctuations of initial energy density and temperature may
survive till the freeze-out and manifest themselves in the
measured temperature fluctuations. 
In fact, initial fluctuating conditions have been found to be necessary for
explaining observed elliptic flow in central collisions and
substantial triangular flow of charged
particles~\cite{hannu}. The initial state fluctuations may have
their imprint on the bin to bin local fluctuations within an event.
Event-by-event hydrodynamic calculations provide
a strong theoretical basis for studying the global and local
temperature fluctuations.

\section{Event-by-Event hydrodynamical approach}

Local fluctuations in energy density arise because of the internal
structures of the colliding nuclei.
These initial fluctuations manifest into local temperature
fluctuations of the fireball at different stages of the collision.
Relativistic hydrodynamic calculations which take such effects into account reveal 
large local fluctuations in $\epsilon$ and $T$ in small phase space
bins at early stages of the collision~\cite{heinz,hannu,song,Urs-1}.
The local fluctuations have been
quantified throughout the evolution by simulating central (0--5\% of the total cross section) 
Pb-Pb events at \sNN~=~2.76 TeV by the 
use of a (2+1)-dimensional event-by-event ideal hydrodynamical
framework~\cite{hannu} with lattice-based equation of state~\cite{eos}.
The formation time of the plasma
is taken to be 0.14~fm~\cite{phe,chre1}.
A wounded nucleon (WN) profile is considered where the initial
entropy density is distributed using a 2-dimensional
Gaussian distribution function,
\begin{eqnarray}
s(X,Y)  =   \frac{K}{2 \pi \sigma^2}  \sum_{i=1}^{\ N_{\rm WN}} 
\exp \Big( -\frac{(X-X_i)^2+(Y-Y_i)^2}{2 \sigma^2} \Big).
\label{eqn1}
\end{eqnarray}
Here $X_i,Y_i$ are the transverse coordinates of the $i^{\rm th}$ nucleon and $K$ is an
overall normalization constant. The size of the density fluctuations
is determined by the free parameter $\sigma$, which is taken to be
0.4~fm~\cite{hannu}. 
The transition temperature from the QGP to the hadronic phase is
chosen to be 170~MeV and the kinetic freeze-out temperature is taken as 160~MeV.
The results for $\epsilon$ and $T$ at different
times ($\tau$) are analyzed for each collision in $X$-$Y$ phase space 
bins in the transverse plane (each bin is chosen to be 0.6~fm$\times$0.6~fm).

Time evolutions of the distributions of $\epsilon$ and $T$ have been
presented for four values of $\tau$ in  Fig.~\ref{fig1} for a single event.
The upper panels (a-d) of the figure show three dimensional view of 
$\epsilon$, whereas the lower panels (e-h)
show corresponding values of $T$ in the
transverse plane. 
At early times, sharp and pronounced peaks in
$\epsilon$ and hotspots in $T$ are observed.
Large bin-to-bin fluctuations observed in $\epsilon$ and
$T$ indicate that the system formed immediately after collision is quite
inhomogeneous in phase space. As time elapses, the system
cools, expands, and the bin-to-bin variations in $\epsilon$ and $T$
smoothen out.

\begin{figure}[tbp]
\centering
\includegraphics[width=0.5\textwidth]{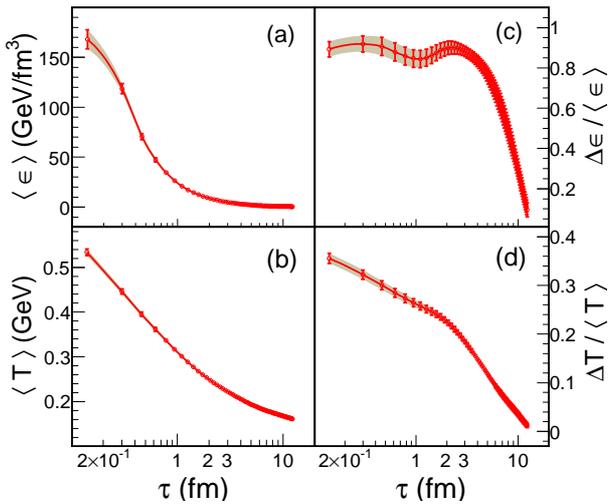} 
\caption{(Color online). Temporal 
evolution of (a) average energy density, (b) average temperature,
(c) fluctuations in energy density, and (d) fluctuations in temperature 
for central Pb-Pb 
collisions at \sNN~=2.76 TeV, obtained from hydrodynamic
calculations. Averaging is taken over the $X$-$Y$ bins in every event.
The shaded regions represent the extent of event-by-event variations
for a large number of events.
}
\label{fig2}
\end{figure}

Observations from Fig.~\ref{fig1} have been quantified in terms of the mean 
energy density (\energy), mean temperature (\temp) over the $X$-$Y$
bins, and the
bin-to-bin fluctuations of $\epsilon$ and {$T$} as a function of $\tau$. 
The time evolution of \energy, \temp, 
and their fluctuations are presented in Fig.~\ref{fig2}, where
$\tau$ is plotted in logarithmic scale for zooming in on the early times. 
The shaded regions represent the extent of event-by-event variations,
taken from a sample of five hundred events.
It is observed that
\energy~falls sharply from $\sim$168~GeV/fm$^3$ at
$\tau=0.14$~fm to a value of $\sim$20~GeV/fm$^3$ at $\tau=1$~fm, and
then falls slowly till freeze-out. 
The initial $\epsilon$ is close to the experimental result,
estimated by the ALICE collaboration~\cite{alice_energy,toia}.
The fall of \temp~with $\tau$ is smooth, which goes down
from $\sim$530~MeV at
$\tau=0.14$~fm to $\sim$300~MeV at $\tau=1$~fm.  
At the freeze-out, as expected, \temp~is close to 160~MeV.

The bin-to-bin fluctuations,
$\Delta \epsilon/\langle \epsilon \rangle$ and $\Delta T/\langle T \rangle$,
are presented in the right panels of Fig.~\ref{fig2},
where $\Delta \epsilon$ and  $\Delta T$ are the root mean
square (RMS) deviations of the two quantities over the bins. 
At early times, the fluctuations 
are observed to be very large ($\sim$90\% and
$\sim$35\% for $\epsilon$ and $T$, respectively),  which indicate
the violent nature of the matter created in the collision. 
Interestingly, although \energy~decreases quite fast, 
the fluctuation in $\epsilon$ remains almost constant up to
$\tau$$\sim$2.5~fm, and then decreases rapidly. Around the same
$\tau$, the fluctuation in $T$ shows a kink, beyond which the
fluctuation decreases even faster.
During the hydrodynamic evolution, 
there may be a characteristic change in the behaviour of the system at
this time, which needs to be understood.
Nevertheless, it is clear that 
a detailed insight to the
evolution of fluctuations is possible by studying local temperature
fluctuations.

\section{Global Temperature Fluctuations}

Fluctuations of any observable of a system have two distinct origins, first, quantum fluctuations
which are initial state fluctuations occurring at fast time scales
and second, classical thermodynamical fluctuations which occur after elapse
of sufficient time after the collision~\cite{landau,kapusta}. 
Initial state fluctuations arise because of internal structures of the
colliding nuclei, and fluctuation in initial energy densities, 
and appear as event-by-event fluctuations in the energy density or temperature.
Thermodynamic fluctuations have multiple sources, such as local
thermal fluctuations of energy density, and event-by-event variation
in the freeze-out conditions. Fluctuations of these observables are
related to several thermodynamic parameters. For example,
fluctuation in temperature are related to the heat
capacity of the system~\cite{Sto,landau}:
\begin{eqnarray}
\frac{1}{C} = \frac{  (\Delta T^2) }  {T^2},
\label{eqn2}
\end{eqnarray}
where $T$ is the temperature of the 
system at freeze-out\footnote{
For part of a thermodynamically equilibrated system with fluctuating
energy ($E$), the heat capacity \cite{steph2} can be expressed through
 $  { (\Delta E^2) } = {T^2}C(T) $. 
This can be applied for a locally thermalized system
produced during the evolution of heavy-ion collisions.
                  }, 
and $\Delta T^2 = \langle T^2 \rangle - \langle T \rangle ^2 $.

Fluctuations in a thermodynamic quantity like temperature may arise
due to fluctuations in energy or entropy or finite particle number,
and can be calculated from  \meanpt~(mean transverse momentum)
fluctuations or \pt~correlations~\cite{shuryak,steph2,wilk1,korus,morw,voloshin,na49,star-PT}. 
However, according to Ref.~\cite{steph2}, variation of \cv~is similar to fluctuation in energy
rather than the \pt~fluctuation. 
Thus \pt~correlation is not an appropriate measure of the temperature
fluctuation, and may also affect measurement of \cv. 
To circumvent this limitation, we have come up with a new approach by
assigning a temperature to an event and estimating event-by-event
fluctuation for a given class of events.
In fact, for central Pb-Pb collisions at the LHC energy, large number
of produced particles in each event form a well 
defined exponential
distribution in \pt. The inverse slope parameter of the distribution
gives the effective temperature, and subsequently,  the heat capacity of the
system can be calculated by using eqn.~(\ref{eqn2}).

A feasibility study of
calculating the global event temperature and local
temperatures in small phase space bins has been made for Pb-Pb
collisions at \sNN~=~2.76 TeV,
corresponding to the LHC energy. 
The string melting (SM) mode of the AMPT model~\cite{ampt} has been
employed, which mimics the experimental conditions.
This mode of AMPT model, 
using parameter set-B as given in Ref.~\cite{cmko1,cmko2},
includes a fully partonic phase that hadronizes through
quark coalescence, and has been shown to reproduce the experimental
data at LHC energy~\cite{subrata,cmko1,cmko2,dronika}.  

\begin{figure}[tbp]
\centering
\includegraphics[width=0.49\textwidth]{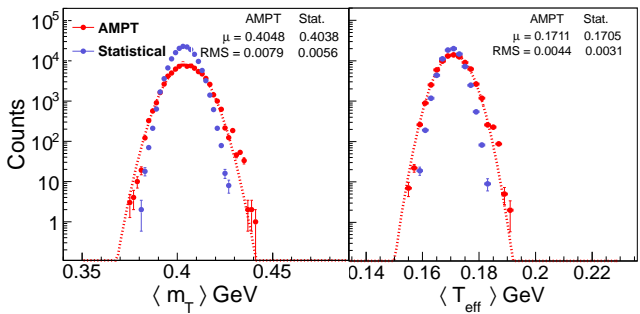} 
\caption{(Color online). 
Event-by-event distribution of \meanmt~and corresponding \Tevt for pions within
one unit of rapidity and full azimuth for central Pb-Pb collisions
at \sNN~=~2.76 TeV using AMPT. Distributions for synthetic events
representing the statistical component are
overlaid on the figure.
}
\label{fig3}
\end{figure}

A large number of particles produced in every central Pb-Pb collisions
at \sNN~=2.76 TeV
makes it possible to construct \mt~spectra of identified
particles in every event. Fitting the \mt~distribution of pions with a 
Maxwell-Boltzmann distribution yields the value of the effective 
temperature, \Tevt in each event.
This temperature is related to the \meanmt~of pions within a given acceptance:
\begin{eqnarray}
\langle m_{\rm T} \rangle = \frac{2T_{\rm eff}^2 + 2m_0 T_{\rm eff} + m_0^2}
{m_0 + T_{\rm eff}},
\label{eqn3}
\end{eqnarray}
where $m_0$ is the mass of pions.
From the AMPT events, it is verified that for pions within the
central rapidity 
($-0.5 \le y \le 0.5$), full azimuth and 
 $m_0 \le m_{\rm T} \le 1.5$~GeV,
the value of \Tevt from fitting the \mt~spectrum as well as from 
\meanmt~are similar to each other.
The upper limit of \mT~is chosen to exclude pions, which may be affected
by jets. 
Figure~\ref{fig3} shows the distributions of \meanmt~of pions and
corresponding values of 
\Tevt calculated using the above expression for a large number of
events. 
It should be noted that the values of \Tevt have
contributions from both a thermal part ($T_{\rm kin}$) and a second component
which depends on the collective transverse 
velocity~($\langle \beta_{\rm T} \rangle$) of the system such that 
$T_{\rm eff} = T_{\rm kin} + f(\beta_{\rm T})$.
For pions, $ f(\beta_{\rm T}) \approx  m_0 \langle \beta_{\rm T}\rangle ^2 $~\cite{jajati-jane}.
For top central (0-5\%) collisions, 
the transverse velocity can be assumed to be same for all events due
to spherical symmetry, and the
fluctuations in the flow velocity can be neglected~\cite{steph2}.
Thus the fluctuation in \Tevt may be a good
representation of the fluctuation in temperature,
($\Delta T_{\rm kin} =  \Delta T_{\rm eff}$). Being a thermodynamic
quantity, the heat capacity is only connected to $T_{\rm kin}$, and
eqn.~\ref{eqn2} can be rewritten as:
\begin{eqnarray}
\frac{1}{C} 
= \frac{  (\Delta T_{\rm kin}^2) } {T_{\rm kin}^2}
\approx \frac{ (\Delta T_{\rm eff}^2)} {T_{\rm kin}^2}.
\label{eqn4}
\end{eqnarray}

Event-by-event global temperature fluctuation 
has contributions from 
both the statistical and dynamical components. The statistical
component arises due to limited number
of particle production in heavy-ion collisions, and does not
contribute to the measurement of heat capacity of the system.  
Statistical component of the fluctuation 
has been estimated by constructing randomly
generated ``synthetic''
events, keeping the multiplicity and \pt~distributions similar to
those of the AMPT events. The 
\meanmt~and \Tevt distributions of the synthetic events are also
shown in Fig.~\ref{fig3}. 
The dynamical component of the
global temperature fluctuation, obtained by subtracting the
statistical component of the fluctuation from the total, has been
estimated to be $\Delta T_{\rm eff} = 3.12 $ MeV. 
According to Ref.~\cite{ALICE_identified}, $T_{\rm kin} = 0.095$ and
$\langle \beta_{\rm T} \rangle = 0.651$ from combined Blast Wave fits.
This corresponds to heat capacity of  $C = 927.1$.  

Specific heat of the system is defined as heat capacity per
volume at the
thermodynamic limit. For ideal gas, specific heat  is  (3/2)N, where N
is the Avogadro's number. For finite number of particles, the specific
heat can be expressed as heat capacity per particle. We define the
specific heat ($c_v$) as the heat capacity per pion multiplicity within
the central rapidity ($-0.5 \le y \le 0.5$) and $0 \le \phi \le \pi$).
The $c_v$ is found to be 0.633. 
Considering the variation of $\beta_T$ within 0.63 and
0.67~\cite{ALICE_identified}, the value of $c_v$
turns out to be within the range of 0.594 to 0.687.
This is the first estimation of $c_v$ at LHC energies from
model calculations.


\section{Local Temperature Fluctuations}

Local temperature fluctuations, which provide the amount of non-uniformity 
within a single event, are studied by dividing the available phase
space into several \yphi~bins, and estimating the bin temperature
(\Tbin). 
It helps to find local hotspots created during the initial energy
density, whether those have survived or died out.
The bin temperatures are obtained using the similar
prescription as above. The \meanmt~of pions are calculated within the \yphi~bin and
$m_0 \le m_{\rm T} \le 1.5$~GeV, and \Tbin~is evaluated by using eqn.~(\ref{eqn3}).
The number of \yphi~bins has been chosen by taking into account
the number of pions in each bin so that fluctuations in the number of
pions do not affect the estimation of \mT. 
The amount of local fluctuation may vary depending on the number of
y–φ bins, which needs to be evaluated.

\begin{figure}[tbp]
\centering
\includegraphics[width=0.49\textwidth]{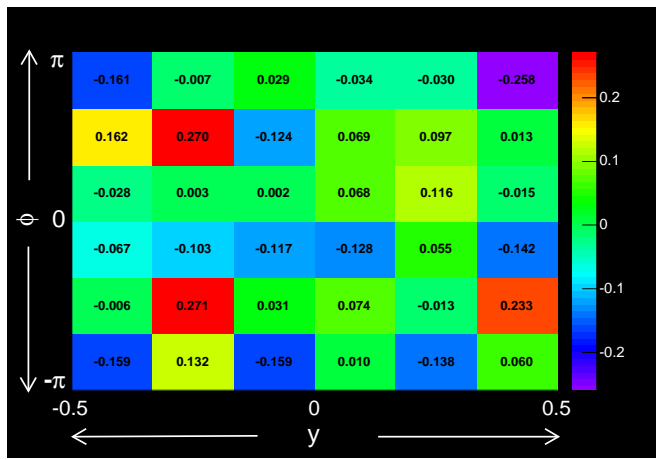} 
\caption{(Color online). 
Temperature fluctuation map in \yphi~bins
for central Pb-Pb collisions at \sNN~=~2.76 TeV using the AMPT model. 
For each \yphi~bin, fluctuation is expressed as 
$(T_{\rm bin} - T_{\rm eff}) / T_{\rm eff}$, 
the deviation of the bin temperature to the event
temperature.
The colour palettes indicate the magnitude of fluctuations.
}
\label{fig4}
\end{figure}

For a given event, local temperature fluctuation in a given \yphi~bin is
expressed as:
\begin{eqnarray}
 F_{\rm bin} = (T_{\rm bin} - T_{\rm eff})/T_{\rm eff}.
\label{eqn5}
\end{eqnarray}
For each event, a fluctuation map in \yphi~phase space
is constructed by plotting the corresponding values of $F_{\rm bin}$.
Fig.~\ref{fig4} shows the the temperature fluctuation map for a
typical event in 6$\times$6 bins in \yphi, where the fluctuations are
represented by different colour pallets.
This map gives a quantitative view of the local temperature fluctuation in
the available phase space for an event.  
The map shows several hot (red) as well as cold (blue) zones, and
zones with average (green) fluctuation throughout the phase space.
It is to be seen whether 
the hot and cold zones have their origin from the extreme
regions of phase space that existed during the early stages of the
reaction.

\begin{figure}[tbp]
\centering
\includegraphics[width=0.49\textwidth]{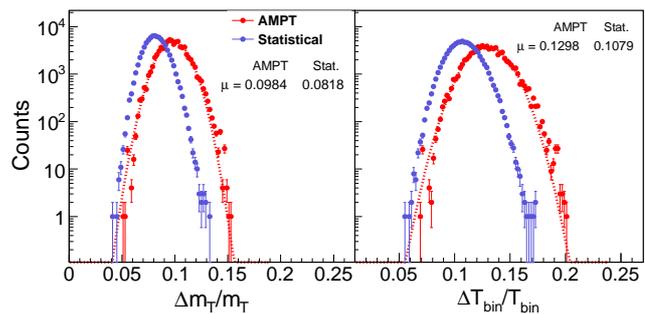} 
\caption{(Color online). 
Event-by-event local temperature fluctuations, obtained from 
$6\times 6$ \yphi~bins in central rapidity and full azimuth for central 
Pb-Pb collisions at \sNN~=~2.76 TeV using the AMPT event generator.
}
\label{fig5}
\end{figure}

For a single event, 
the amount of local temperature fluctuation is
quantified by the ratio of RMS to the mean of the \Tbin~distribution.
Local fluctuations in \meanmt~and corresponding temperature for each event
have been evaluated and their event-by-event distributions are plotted
in Fig.~\ref{fig5}. The left and right panels of the figure show the
distributions from \meanmt~and \Tbin, respectively. 
Mean value of the local temperature fluctuation for the event sample
is 12.98\% for 6$\times$6 bins in central rapidity. Statistical
component of the local fluctuations has been extracted from the
``synthetic'' events as discussed earlier. The mean value
corresponding to the statistical component of the local temperature
fluctuation has been estimated to be 10.79\% as shown in 
Fig.~\ref{fig5}. The average dynamical local fluctuation, extracted
after subtracting the statistical component, for AMPT events is 
7.2\%. 
The non-zero value of local temperature fluctuation may imply that these might
have the contributions from the early state fluctuations.

\section{Remarks}

Extraction of temperature fluctuations from experimental data maybe
affected by some of the effects.
Event plane orientation is one such effects, which needs to be taken
care of, especially for non-central events. For the present study using
AMPT, the events are event plane oriented. 
Fourier decomposition of the momentum distributions in the transverse plane 
yields a $\phi$--independent, axially symmetric radial flow component and a
$\phi$--dependent part containing the anisotropic flow coefficients. 
For most central collisions,
radial flow remains similar for all the events and the anisotropic
flow components do not affect the slope of the \mT~distribution. 
Final state effects, such as resonance decay, and hadronic rescattering
tend to make the \mT~spectra softer, and so choice of the \mT~window has
to be made in order to minimize such effects. Although present
analysis uses charged pions,
species dependence of temperature fluctuations may provide extra
information regarding their freeze-out hyper surfaces as the particle
production mechanisms of mesons, baryons and strange particles are
different. This study may shed light on whether the origin of the 
temperature fluctuations are solely due to initial state
fluctuations or includes final state effect.
Viscosity tends to dilute the fluctuations. 
The SM version of AMPT includes the effect of viscosity 
($\eta/s \sim 0.15$ at $T$=436~MeV~\cite{subrata,dronika}).
Further analysis using a viscous hydrodynamic models will be more
realistic for this study.

Local temperature fluctuation map for each event, as shown in
Fig.~\ref{fig4}, has a striking similarity to the fluctuation map of the
cosmic microwave background radiation (CMBR)~\cite{cmbr1}.
Fluctuation analysis of CMBR fluctuation map  confirms the Big Bang evolution,
inflation and provides information regarding the early Universe. 
The study of higher order moments using the maps may give access to
various thermodynamic parameters at the early stages of the
evolving system. Similarly, the fluctuation maps of heavy-ion
collisions may form the basis of power spectrum analysis~\cite{ajit,paul,morphology}. 
Access to large number of events in heavy-ion collisions compared to
single event analysis in CMBR may have definite advantage which 
can be utilized to our advantage in order
to gain access to conditions that prevailed at the primordial state. 

\section{Summary}

Event-by-event temperature fluctuations 
over full phase space as well as local phase
space bins have been proposed to characterize 
the hot and dense system produced in heavy-ion collisions at
ultra-relativistic energies. The global temperature fluctuations
provide the heat capacity as well as specific heat of the system, whereas the observation of local 
fluctuations would imply the presence of fluctuations at early
stages of the collision. 
Relativistic hydrodynamic calculations have been used to understand
the evolution of $\epsilon$ and $T$ fluctuations. It shows
that the system exhibits fiercely large fluctuations at early times,
which diminishes with the elapse of time.

The feasibility of studying temperature fluctuations in  
Pb-Pb collisions at \sNN=2.76 TeV has been demonstrated by using
simulated events from the AMPT model.
Temperatures are extracted from \meanmt~of charged pions. The global
fluctuation in the event temperature 
has been extracted and used to calculate the heat capacity as $C =
927.1$. The specific heat per pion has been extracted to be $c_v =
0.633 \pm 0.054$ at the LHC energy, which is consistent with the
results at RHIC energies~\cite{chinese}.
At the LHC, the phase transition
is expected to be a cross over. Thus the transition may not take
place at a unique point in the phase diagram due to the evolution of
entropy fluctuation created at the initial state, which reflects into 
energy density and temperature fluctuations. Thus, the estimation of the heat
capacity helps in understanding the nature of the phase transition.
The variation of $c_v$ as a function of center-of-mass beam
energy for Au+Au collisions at RHIC may provide an effective tool for
locating the QCD critical point~\cite{star_CP,gavai_CP}.
At LHC energies, it is possible to extract local temperatures over
small phase space bins in central rapidity. Extraction of Local temperature
fluctuation complements the global event temperature fluctuation as
the origin of local fluctuations may be of primordial in nature. 
For the AMPT, local temperature fluctuations over small phase bins within each event 
have been extracted. The amount of local fluctuation is 7.2\% for
 6$\times$6 bins in central rapidity. The observation of the non-zero local
fluctuation may imply that a part of the initial fluctuations might have
survived till freeze-out. The present
study of global and local temperature fluctuations in conjunction with theoretical model calculations
open up new avenues for characterizing the heavy-ion collisions.

\begin{acknowledgements}
We would like to thank H. Holopainen for providing us with the event-by-event 
hydrodynamic code. 
This research used resources of the LHC grid computing facility at the
Variable Energy Cyclotron Centre. SB is supported by the Department of
Atomic Energy, Government of India. 
\end{acknowledgements}


\end{document}